\documentclass{emulateapj}
\usepackage{xspace}
\usepackage{amsmath}
\usepackage{hyperref}
\usepackage{framed} 
\usepackage{txfonts}
\usepackage{epstopdf} 
\def\euv{\epsilon_{\rm UV}}
\def\Msun{{\rm M}_\odot}

\def\ion#1#2{\textsc{#1#2}}
\def\HI{{\ion{H}{I}}}

\def\H2{{{\rm H}_2}}
\def\lya{Ly$\alpha$}
\def\dew{D_{\rm EW}}

\def\dim#1{\mbox{\,#1}}
\def\hide#1{}

\begin{document}

\title{Cosmic Reionization On Computers II. Reionization History and Its Back Reaction on Early Galaxies}

\author{Nickolay Y.\ Gnedin\altaffilmark{1,2,3} and Alexander A.\ Kaurov\altaffilmark{2}}
\altaffiltext{1}{Particle Astrophysics Center, 
Fermi National Accelerator Laboratory, Batavia, IL 60510, USA; gnedin@fnal.gov}
\altaffiltext{2}{Department of Astronomy \& Astrophysics, The
  University of Chicago, Chicago, IL 60637 USA; kaurov@uchicago.edu}
\altaffiltext{3}{Kavli Institute for Cosmological Physics, The University of Chicago, Chicago, IL 60637 USA} 

\begin{abstract}
We compare the results from several sets of cosmological simulations of cosmic reionization, produced under Cosmic Reionization On Computers (CROC) project, with existing observational data on the high-redshift \lya\ forest and the abundance of \lya\ emitters. We find good consistency with the observational measurements and the previous simulation work. By virtue of having several independent realizations for each set of numerical parameters, we are able to explore the effect of cosmic variance on observable quantities. One unexpected conclusion we are forced into is that cosmic variance is unusually large at $z>6$, with both our simulations and, most likely, observational measurements are still not fully converged for even such basic quantities as the average Gunn-Peterson optical depth or the volume-weighted neutral fraction. We also find that reionization has little effect on the early galaxies or on global cosmic star formation history, because galaxies whose gas content is affected by photoionization contain no molecular (i.e.\ star-forming) gas in the first place. In particular, measurements of the faint end of the galaxy luminosity function by JWST are unlikely to provide a useful constraint on reionization. 
\end{abstract}

\keywords{cosmology: theory -- cosmology: large-scale structure of universe --
galaxies: formation -- galaxies: intergalactic medium -- methods: numerical}

\section{Introduction}
\label{sec:intro}

If cosmic reionization can be called the current frontier of extragalactic astronomy, then, in historic terms, we live in the middle of XIX century. I.e., the frontier is being settled...

Ultra Deep Field campaigns with the Hubble Space Telescope pushed the search for the most likely reionization sources - young star-forming galaxies - to double digit values of cosmic redshift \citep{gals:biff07, gals:biol11,rei:obig12,rei:btos12,rei:sreo13,rei:wmhb13,rei:obil13,rei:bdm14,rei:obi14}. Observations of \lya\ emitters at $z\sim7$ \citep{lae:hcb10,lae:osf10,lae:pfv11,lae:ksm11,lae:sse12,lae:oom12,lae:cbw12,lae:cbw14} indicate rapid change in their abundance as one rides deeper into the frontier territory. Recent mind-blowing progress of the first generation experiments for detecting the redshifted 21cm signal from the epoch of reionization \citep{rei:pla13,rei:dlw13} promises a major observational breakthrough well before the end of this decade. Even along the well-trodden ``Oregon Trail'' of \lya\ absorption spectroscopy of high redshift quasars new advances are expected in the nearest future, as new discoveries of $z>6$ quasars continue \citep{rei:bvm13,rei:vfs14}.

Theoretical studies did not stay behind the observational strides, rejuvenating a somewhat slowed-down progress of the second half of the last decade. The major push on the theory side was galvanized by the pioneering idea of \citet{reisam:fhz04}, who realized that the standard lore of large-scale structure theory, Excursion Set formalism, can be applied to studying the reionization process. That idea generated a large following of semi-analytical and semi-numerical approaches for modeling reionization \citep{reisam:fo05,reisam:fmh06,reisam:mf07,reisam:aa07,reisam:zmm11,reisam:mfc11,reisam:aa12,reisam:zgl13,reisam:btc13,ng:kg13,reisam:sm14}, and more traditional models were pursued as well \citep{reisam:cf05,reisam:cf06,reisam:sv08,reisam:mcf11,reisam:vb11,reisam:mcf12,reisam:kf12,reisam:rfs13}. Unfortunately, on the numerical simulation front the progress was less dramatic, although important advances in the simulation technology did take place \citep[e.g.][for a complete review see \protect\citet{ng:gt09}]{rei:imp06,rei:zlmd07,rei:mlz07,rei:tcl08,rei:stc08,rei:ca08,rei:lcg08,rei:ipm09,rei:at10,rei:fma11,rei:ais12,rei:sim12}. However, the primary brake on the simulation progress - insufficient computing power - is finally being released, thanks to Moore's Law.  

Modern High Performance Computing platforms have crossed an important threshold of ``sustained peta-scale'' performance. This level of performance, currently available on about a dozen or so (non-classified) supercomputers across the globe, offers a unique opportunity for reionization theorists to make a substantial breakthrough in our ability to model cosmic reionization with high physical fidelity, and some of the most recent simulation work already took advantage of that opportunity \citep{rei:ima14,newrei:snr14,newrei:nrs14,newrei:hdp14}. 

Cosmic Reionization On Computers (CROC) project is another effort in producing peta-scale simulations of reionization in sufficiently large volumes (above $100\dim{Mpc}$ in comoving units), with spatial resolution reaching down to $100\dim{pc}$, and including most (if not all) of the relevant physical processes, from star formation and feedback to radiative transfer.

In the first paper in the series \citep[][hereafter Paper I]{ng:g14a} we described in complete detail the simulation design and the calibration of numerical parameters. In this paper we explore the overall process of cosmic reionization as captured by CROC simulations, and compare our theoretical predictions to several observational constraints. 

We deliberately limit the scope of this paper to relatively easily computable quantities, which give only a global, broad-brush view of reionization, due to the limited human effort available for the analysis of the rich, but complex simulation data. We intend to continue this paper series as more detailed, labor-intensive analysis gets completed.

\section{CROC Simulations}
\label{sec:croc}

All CROC simulations are performed with the Adaptive Refinement Tree (ART) code \citep{misc:k99,misc:kkh02,sims:rzk08}. The ART code is capable of modeling a diverse set of physical processes, from the dynamics of dark matter and gas to star formation, stellar feedback, and radiative transfer. A detailed description of all physical processes followed in the CROC simulations is presented in Paper I.

CROC simulations performed in volumes with $20h^{-1}\dim{Mpc}$ and $40h^{-1}\dim{Mpc}$ on a side, and $80h^{-1}\dim{Mpc}$ boxes will be added to the full data set as the project progresses. All simulations but one have the same mass resolution of $7\times10^6\Msun$ ($20h^{-1}\dim{Mpc}$ boxes use $512^3$ dark matter particles, $40h^{-1}\dim{Mpc}$ use $1024^3$ particles, etc). One of $20h^{-1}\dim{Mpc}$ boxes (B20HR.uv2) has been run with $1024^3$ particles, achieving the 8 times higher mass resolution of $9\times10^5\Msun$; we use that simulation for testing numerical convergence and for some of the scientific results where high mass resolution is required. Using the full Adaptive Mesh Refinement (AMR) functionality of ART, we reach formal spatial resolution (smallest simulation cell size) of $125\dim{pc}$ at $z=6$, and even higher resolution at earlier times, as our resolution remains constant in comoving units. The real resolution of the simulations is a factor of 2-3 worse.

Paper I describes numerical parameters of CROC simulations, and how we calibrate their values. The only parameter that we vary in this paper is the ``escape fraction of ionizing radiation up to the simulation resolution'' $\euv$. In a numerical simulation with finite spatial resolution not all absorptions of ionizing photons can be accounted for, because some of the photons will be lost in structures that are not resolved in the simulation (like a parent molecular cloud). Hence, to account for those absorptions, we assign each stellar particle ionizing luminosity
\[
  L_{\rm ion} = \euv L^{\rm orig}_{\rm ion},
\]
where $L^{\rm orig}_{\rm ion}$ is unattenuated luminosity of a single-age stellar population and the parameter $\euv$ accounts for unresolved photon losses.

Since the ionizing output of our model galaxies is proportional to $\euv$, that parameter critically controls the whole process of reionization in the intergalactic medium (IGM).

\begin{table}[b]
\caption{Simulation Sets\label{tab:sets}}
\centering
\begin{tabular}{llll}
\hline\hline\\
Set Id & $\euv$ & Stopping & Number of \\
       &        & redshift & realizations \\
\\
\hline\\
\underline{$20h^{-1}\dim{Mpc}$ boxes, $512^3$ particles} \\
\\
~~~~B20.uv1   & 0.1 & 5 & 6 [A-F] \\
~~~~B20.uv2   & 0.2 & 5 & 6 [A-F] \\
~~~~B20.uv4   & 0.4 & 5 & 3 [D-F] \\
\\
\underline{$20h^{-1}\dim{Mpc}$ boxes, $1024^3$ particles} \\
\\
~~~~B20HR.uv2 & 0.2 & 5.7 & 1 [B]\\
\\
\underline{$40h^{-1}\dim{Mpc}$ boxes, $1024^3$ particles} \\
\\
~~~~B40.uv1 & 0.1 & 5 & 3 [A-C] \\
~~~~B40.uv2 & 0.2 & 5.5 & 3 [A-C] \\
\\
\hline
\\
\\
\end{tabular}
\end{table}

For each value of simulation parameters (box size, $\euv$, etc) we perform a set of simulations that start from independent realizations of initial conditions and properly account for the fluctuations outside the box using the so-called ``DC mode'' \citep{ng:gkr11}. Hence, we can use a given simulation set to quantify the effect of cosmic variance on our results.

Table \ref{tab:sets} lists simulation sets that we use in this paper. Star formation and stellar feedback parameters in these simulations are calibrated so that the observed galaxy UV luminosity functions are matched to the observations at all redshifts from $z=6$ to $z=10$. Hence, for all simulation sets used in this paper stellar sources of cosmic reionization are followed accurately (at least for $z\leq 10$).

\section{Results}
\label{sec:res}

\subsection{Reionization History}
\label{sec:igm}

\begin{figure}[t]
\includegraphics[width=\hsize]{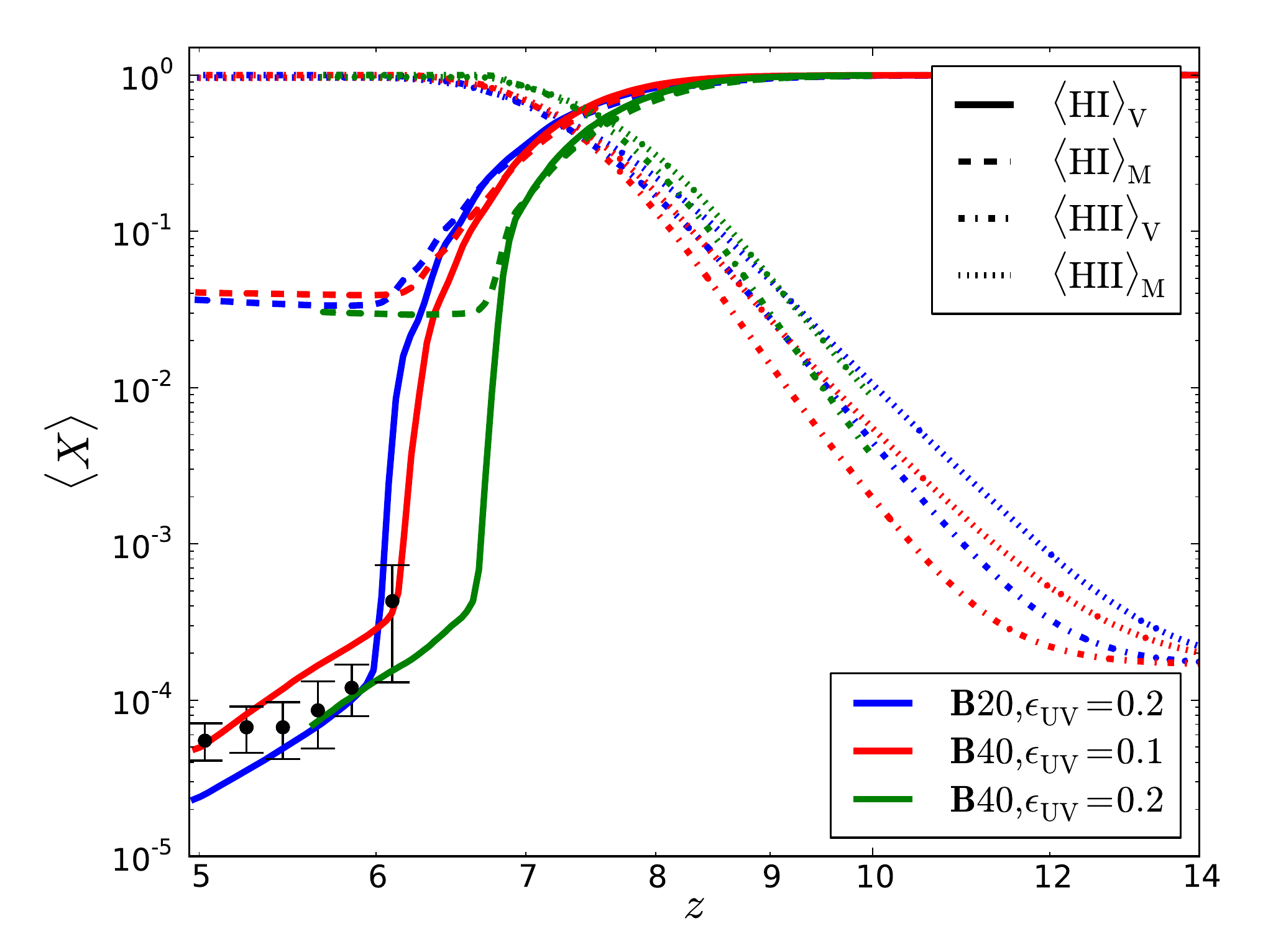}
\caption{Evolution of mass- and volume-weighted hydrogen fractions with redshift in the best-fit $20h^{-1}\dim{Mpc}$ simulation set B20.uv2 and both $40h^{-1}\dim{Mpc}$ sets. Data points are from \protect\citet{igm:fsbw06}.\label{fig:xhz}}
\end{figure}

In Paper I it was showed that the value of $\euv$ between 0.1 and 0.2 provides the best match to the observed evolution of the \lya\ forest at $z<6$ (with 0.2 giving a better fit). This is again illustrated in Figure \ref{fig:xhz}, where we show the evolution of average mass- and volume-weighted hydrogen fractions for the best fit $20h^{-1}\dim{Mpc}$ set (B20.uv2) and both $40h^{-1}\dim{Mpc}$ sets (in all cases we average over all independent realizations). 

Shapes of various curves in the Figure are highly expected, and are consistent between almost all previous simulations of reionization. Ionized fractions grow steadily with time, as ionized bubbles expand in the neutral IGM. Neutral fractions decrease in response until the moment of overlap, when the volume-weighted neutral hydrogen fraction decreases rapidly \citep{ng:g00a}. In the post-overlap stage the IGM is highly ionized, and the evolution of neutral fractions is governed by the mean free path of ionizing photons and the level of cosmic ionizing background \citep[see also][for a general overview of reionization process]{ng:gt09}.

Lack of numerical convergence that we discussed in Paper I is also visible in Figure \ref{fig:xhz} - in the set B40.uv1 the overlap of ionized bubbles (indicated by the rapid drop in the average volume-weighted $\HI$ fraction just before $z=6$) occurs at about the same time as in the smaller box set B20.uv2, but the post-reionization \lya\ forest is better matched by the set B40.uv2, which has a significantly earlier overlap.

As we discussed in Paper I, that lack of convergence is caused by cosmic variance\footnote{Under the term ``cosmic variance'' we understand the difference between separate regions of the universe; such difference is caused both by the variation of densities and by variation in the distribution of ionized bubbles, and the latter almost always dominates.} - having multiple independent realizations allows to explore it well. Our (still unconverged) simulations sample a much larger number of independent sightlines than is available observationally, hence the lack of agreement between the simulations and observations at $z>6$ cannot yet be taken seriously.

At $z<6$ the situation is, however, completely different: simulations with the same value of the $\euv$ parameter do converge, the cosmic variance is small (since the radiation field is dominated by the cosmic background - this is apparent from Figure 4 of Paper I), and the correct simulations should match the observational data. The mismatch between the simulations and the observations at $z\la5.3$ is, therefore, real. In Paper I we also showed that, similarly, our simulations fail to match the galaxy UV luminosity function at $z=5$ well enough. Both these discrepancies indicate that our simulations become inaccurate after $z\approx 5.3$, most likely because our spatial resolution, being kept fixed in comoving units, degrades too much by $z\approx5$.

\begin{figure}[t]
\includegraphics[width=\hsize]{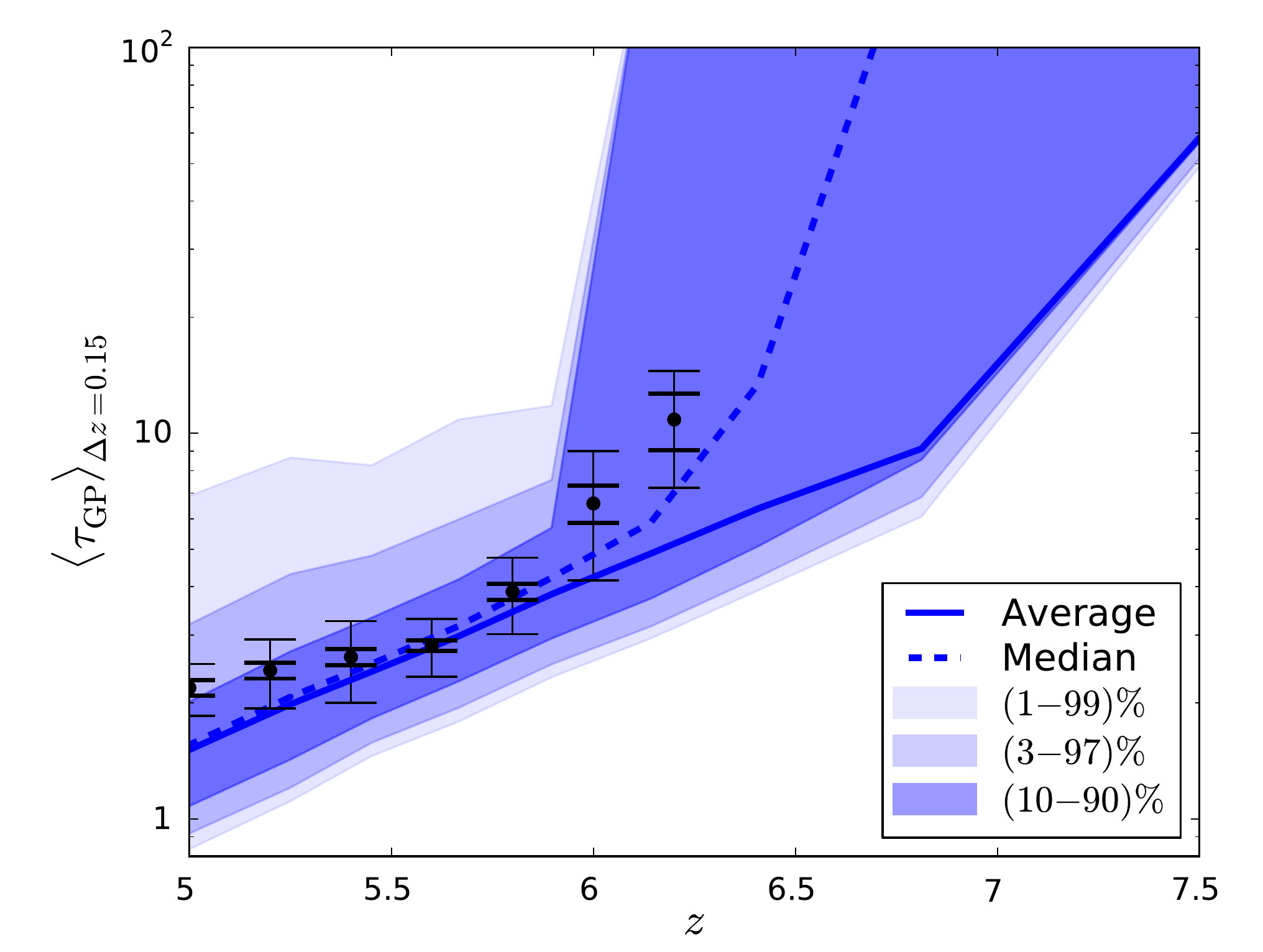}
\caption{Probability Density Function (PDF) of the average Gunn-Peterson optical depth $\langle\tau_{\rm GP}\rangle_{\Delta z=0.15}$ in redshifts intervals $\Delta z=0.15$ along individual lines of sight as a function of redshift for the B20.uv2 simulation set. PDF is shown with 3 progressively more opaque bands that mark progressively narrower percentile ranges around the median (the dashed line). A solid line shows the average, that at $z>6.7$ falls outside (10-90)\% percentile range due to the extreme non-linearity in the relationship between $\tau_{\rm GP}$ and the gas properties. Data points are from \protect\citet{igm:fsbw06}.\label{fig:poftau}}
\end{figure}

Another illustration of the role of cosmic variance is shown in Figure \ref{fig:poftau}, where we plot the distribution of the average Gunn-Peterson optical depth $\langle\tau_{\rm GP}\rangle_{\Delta z=0.15}$ as a function of redshift for individual lines of sight. Lines of sight are generated randomly (i.e.\ starting at random locations and going in random directions) throughout each of the simulation boxes, and synthetic \lya\ spectra are generated along each of them. The optical depth along each line of sight is computed as an average over a distance of $40h^{-1}\dim{Mpc}$ ($\Delta z\approx0.15$), which is well-matched to the redshift interval used in observational studies \citep{igm:fsbw06}. At $z>6$ the distribution becomes exceptionally wide, and not just in the tails. At lower redshift, past overlap, the distribution narrows significantly, but still maintains a relatively long tail towards large values of $\langle\tau_{\rm GP}\rangle_{\Delta z=0.15}$. Hence, we predict that, in a small fraction of all quasar sightlines, segments with $\langle\tau_{\rm GP}\rangle_{\Delta z=0.15}$ as high as 10 can be observed all the way down to $z\approx5.5$.

\begin{figure}[t]
\includegraphics[width=\hsize]{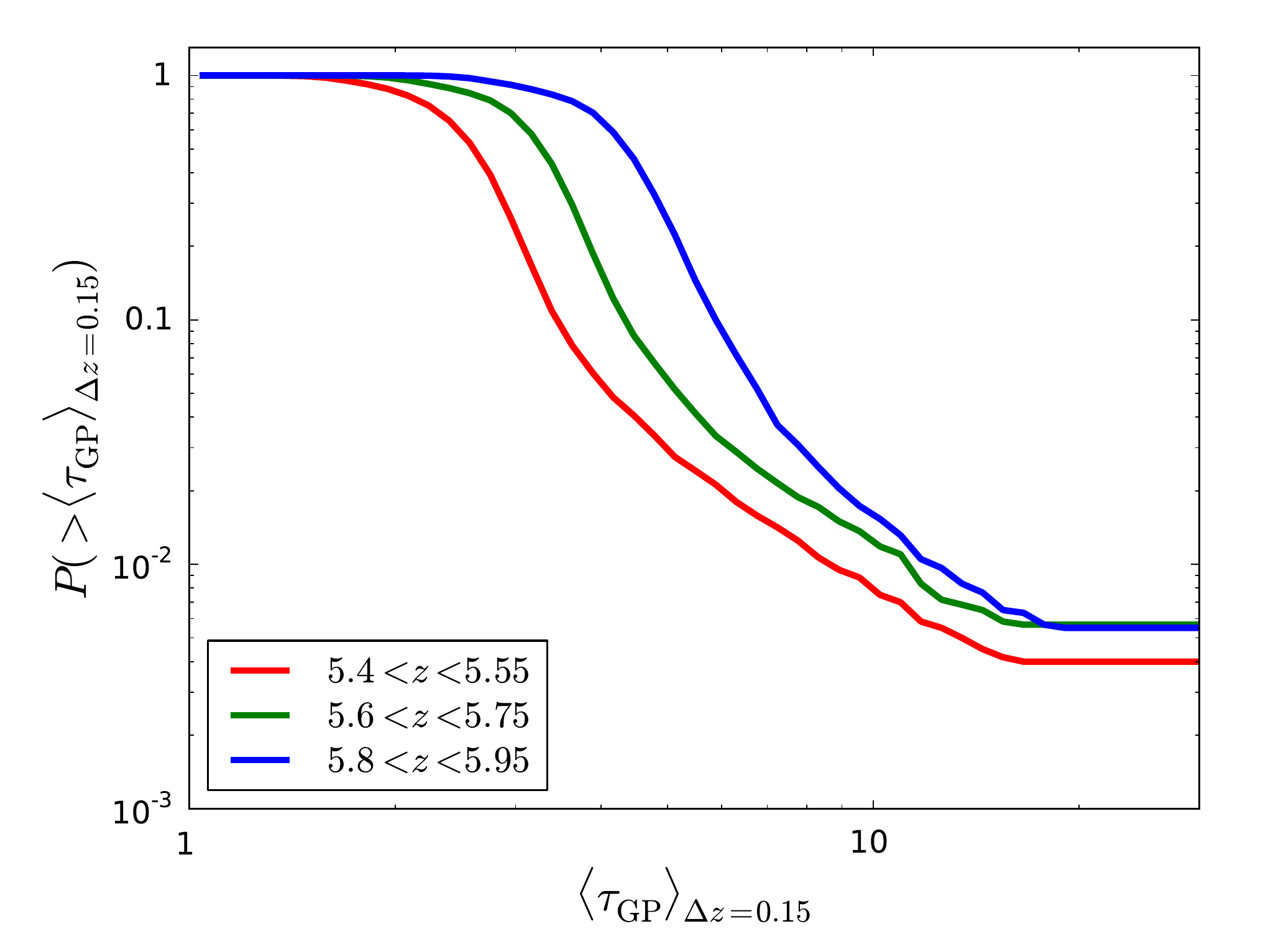}
\caption{Cumulative probability to find a line of sight with the average Gunn-Peterson optical depth $\langle\tau_{\rm GP}\rangle_{\Delta z=0.15}$ above a given value in three redshift bins.\label{fig:poftau2}}
\end{figure}

Figure \ref{fig:poftau2} shows the same result in a more conventional way, where we plot a cumulative probability distribution for $\langle\tau_{\rm GP}\rangle_{\Delta z=0.15}$ in 3 redshift bins after reionization (such a distribution is, obviously, dependent on the redshift interval over which averaging is done). Even at $z\sim 5.5$ there is a 0.4\% probability to find a line of sight with no detectable \lya\ flux. For such a large redshift interval to be devoid of any flux, it is not enough to just have a damped \lya\ in that line of sight, rather it is a genuine feature of the cosmic variance due to large-scale density fluctuations.

\subsection{Ionized Bubbles}
\label{sec:bubbles}

\begin{figure}[t]
\includegraphics[width=\hsize]{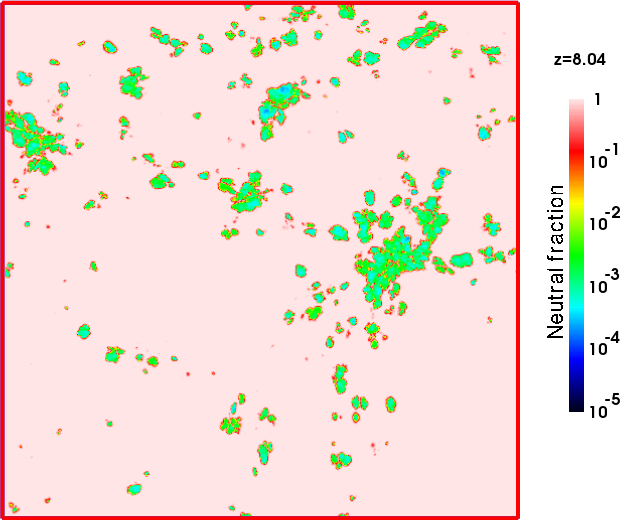}\newline%
\includegraphics[width=\hsize]{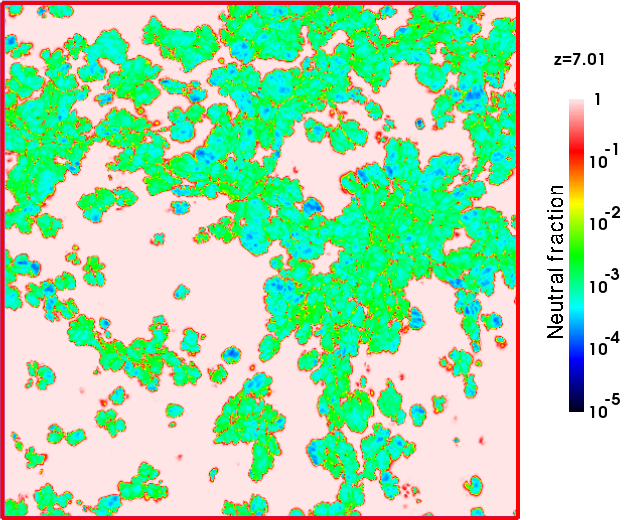}%
\caption{Slices through the computational domain for the first realization of B40.uv1 set (run B40.uv1.A in the notation of Paper I) at $z=8$ and $z=7$. While at $z=8$ individual ionized bubbles can still be identified reasonably well, by $z=7$ the concept of a ``bubble'' becomes ill-defined.\label{fig:slice}}
\end{figure}

Time evolution of the distribution of ionized bubbles is one of the most important characteristics of the reionization process. Unfortunately, in a realistic cosmological simulation the concept of an ``ionized bubble'' is not well defined mathematically, especially at late times, as can be easily seen from Figure \ref{fig:slice}. Hence, in order to have a working definition that can also be compared with other studies, we closely follow the procedure from \citet{rei:zlmd07} to compute the probability that a given point in the simulation is located inside an ionized bubble of size $R$. The scale $R$ is defined as the largest radius of a sphere in which the volume-weighted average ionized fraction is higher than the threshold value, which is chosen to be 90\%. The only difference with \citet{rei:zlmd07} is normalization: we normalize the bubble size distribution to the total volume (an integral over the distribution is equal to the volume-weighted average ionized fraction), whereas \citet{rei:zlmd07} normalize their distributions to the total ionized volume (an integral over the distribution is equal to unity).

\begin{figure}[t]
\includegraphics[width=\hsize]{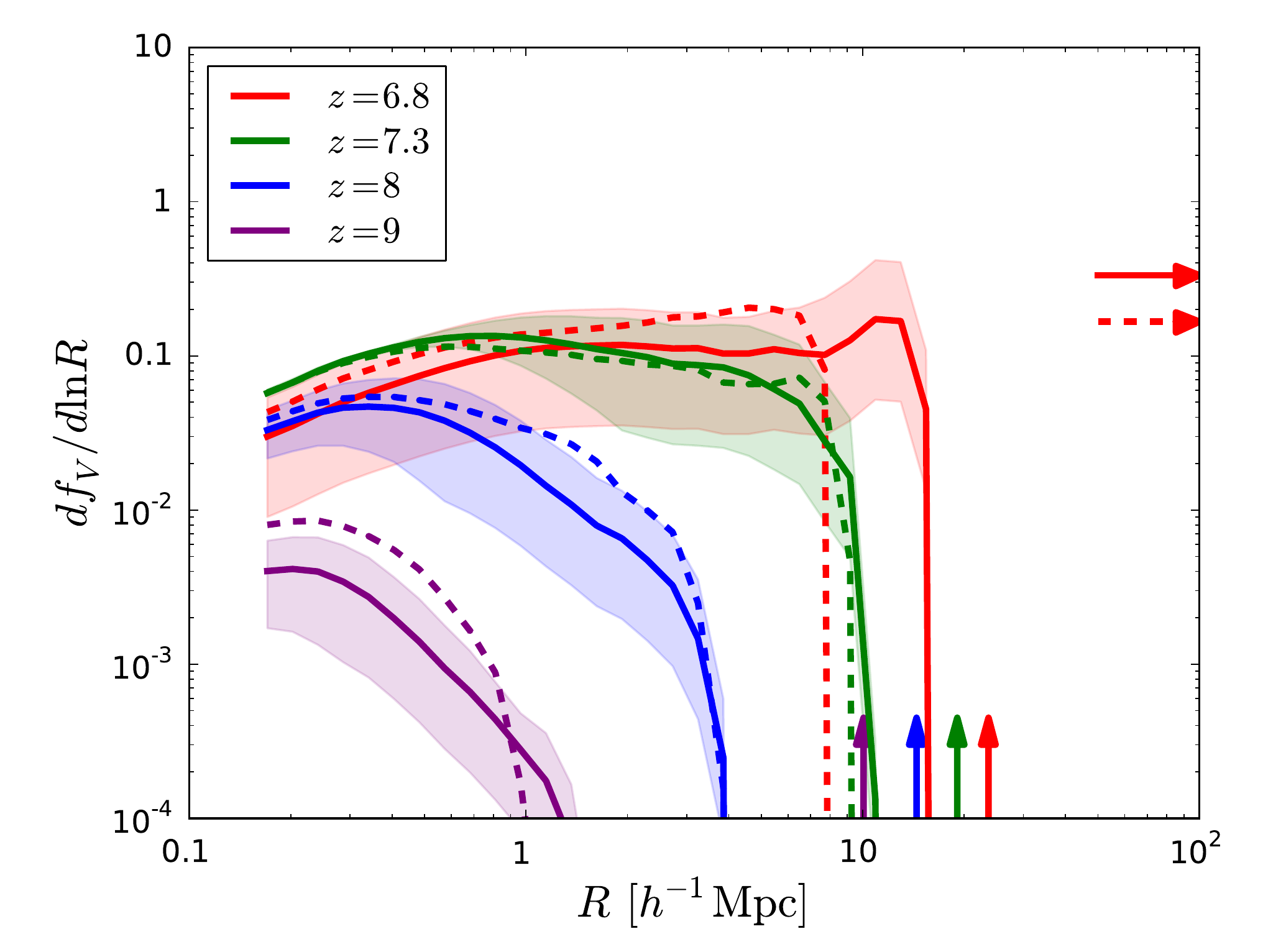}
\includegraphics[width=\hsize]{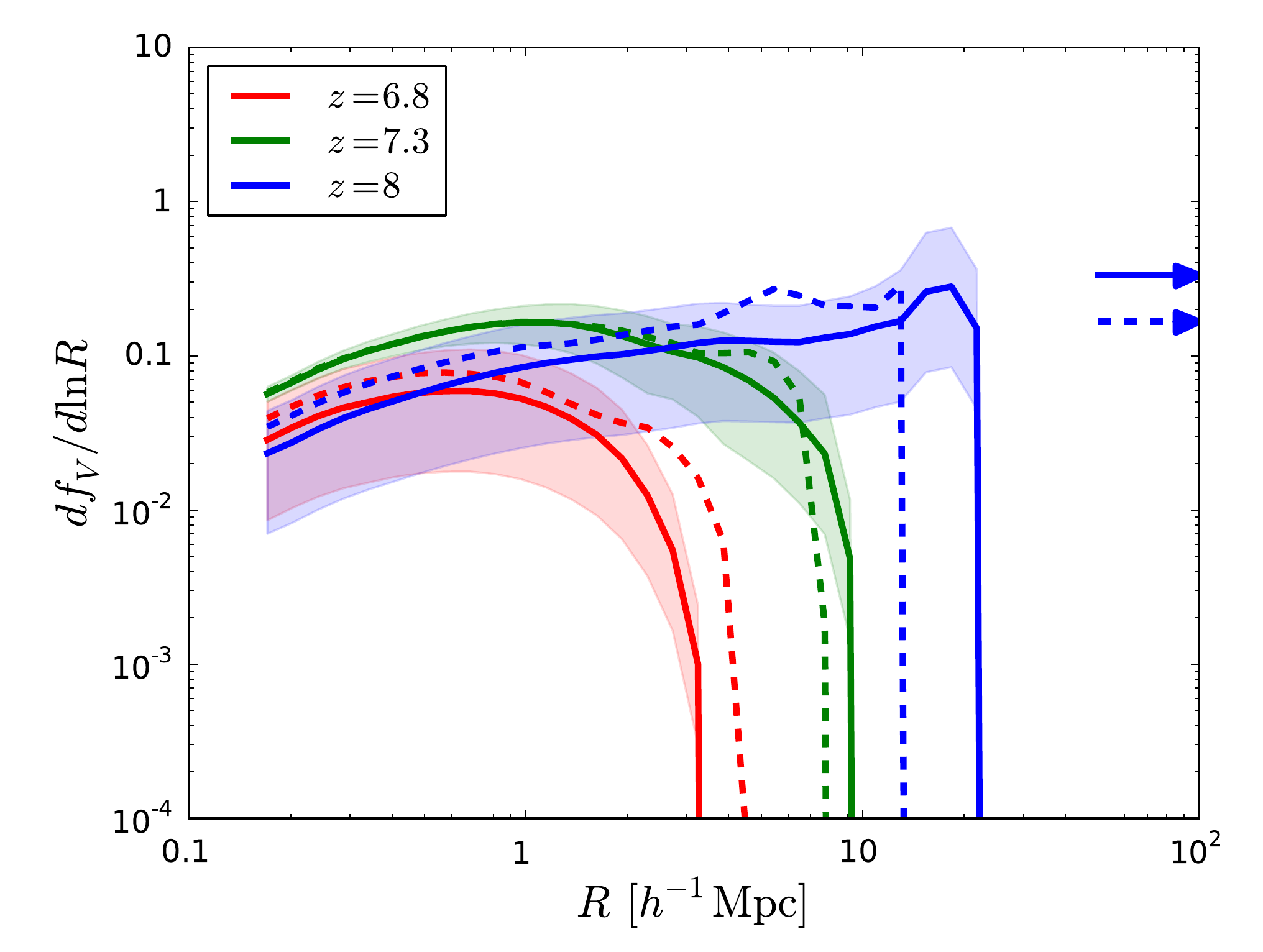}
\caption{Differential volume function of ionized (top) and neutral (bottom) bubbles in our simulations at several values of redshift. Solid lines with semi-transparent bands show the average and the rms for our fiducial $40h^{-1}\dim{Mpc}$ set B40.uv1; dashed lines show averages for the $20h^{-1}\dim{Mpc}$ set B20.uv2. Horizontal errors show the volume fraction in individual boxes that are more than 90\% ionized (top panel) or more than 90\% neutral (bottom panel) at a given redshift. We find good convergence of sizes of ionized bubbles in boxes as small as $20h^{-1}\dim{Mpc}$ at $z>7$, and at $z<9$ for neutral bubbles. For comparison, vertical arrows in the top panel mark the values of the mean free path of ionizing photons due to Lyman Limit system at the corresponding redshifts.\label{fig:bubbles}}
\end{figure}

Distributions of sizes of ionized and neutral bubbles (shown as a differential volume function) are presented in Figure \ref{fig:bubbles} for several values of redshift for both $20h^{-1}\dim{Mpc}$ and $40h^{-1}\dim{Mpc}$ simulation sets. Somewhat unexpectedly, we find good convergence in the bubble size distribution even for the $20h^{-1}\dim{Mpc}$ box size all the way to $z\sim7$. At lower redshifts the convergence does break down, simply because some of the smaller boxes are going to be completely ionized before larger boxes (and the same applies to neutral ``bubbles'' at high redshifts). The volume fraction in such boxes is shown with horizontal arrows on both panels, and it also is reasonably consistent between the $20h^{-1}\dim{Mpc}$ and $40h^{-1}\dim{Mpc}$ simulation sets.

At face value, this result is inconsistent with the recent simulations of \citet{rei:ima14}, who found incomplete convergence in simulation volumes as large as $114h^{-1}\dim{Mpc}$ at all redshifts. Without detailed comparison, it is difficult to isolate the source of the discrepancy. We notice, however, that the spatial resolution of the radiative transfer solver in \citet{rei:ima14} simulations is only about $200h^{-1}\dim{kpc}$ in comoving units, which is inadequate for resolving absorptions in galactic halos and Lyman Limit systems, while our spatial resolution ($0.6h^{-1}$ comoving kpc) is better suited for properly accounting for all absorptions of ionizing radiation.

\subsection{Damping Wing of \lya\ Absorption}
\label{sec:dwing}

Before the universe is completely reionized, patches of the still neutral IGM can significantly absorb \lya\ emission from high-redshift galaxies, as the damping wing of \lya\ absorption extends far redward of the galaxy systemic velocity \citep{igm:m98}. In order to model that effect, we also generate synthetic \lya\ spectra that originate at galaxy locations. The ``sky'' of each galaxy is sampled uniformly with 12 directions, corresponding to 12 zero level cells of the HEALPix\footnote{{\tt http://healpix.sourceforge.net}} tessellation of a sphere \citep{misc:ghb05}. In order to exclude the local absorption from the galactic ISM, we start the line of sight $10\dim{kpc}$ away from the center of the galaxy.

An exact calculation of the effect of the damping wing on the \lya\ emission line of a galaxy requires complex \lya\ radiative transfer in the galactic ISM and surrounding IGM. Such a calculation is a separate research project in itself, and in any case our simulations do not have enough spatial resolution to perform such a calculation with sufficient accuracy. Instead, we approximate the effect of the damping wing by computing the absorption equivalent width of the red part of the synthetic spectrum,
\[
  \dew = \int_{\lambda_0}^\infty e^{\displaystyle-\tau_\lambda} d\lambda,
\]
where $\lambda_0$ is the wavelength of \lya\ at the systemic velocity of each model galaxy. Hence, we compute 12 values of $\dew$ for each galaxy, achieving dense sampling of the full distribution function for $\dew$.

\begin{figure}[t]
\includegraphics[width=\hsize]{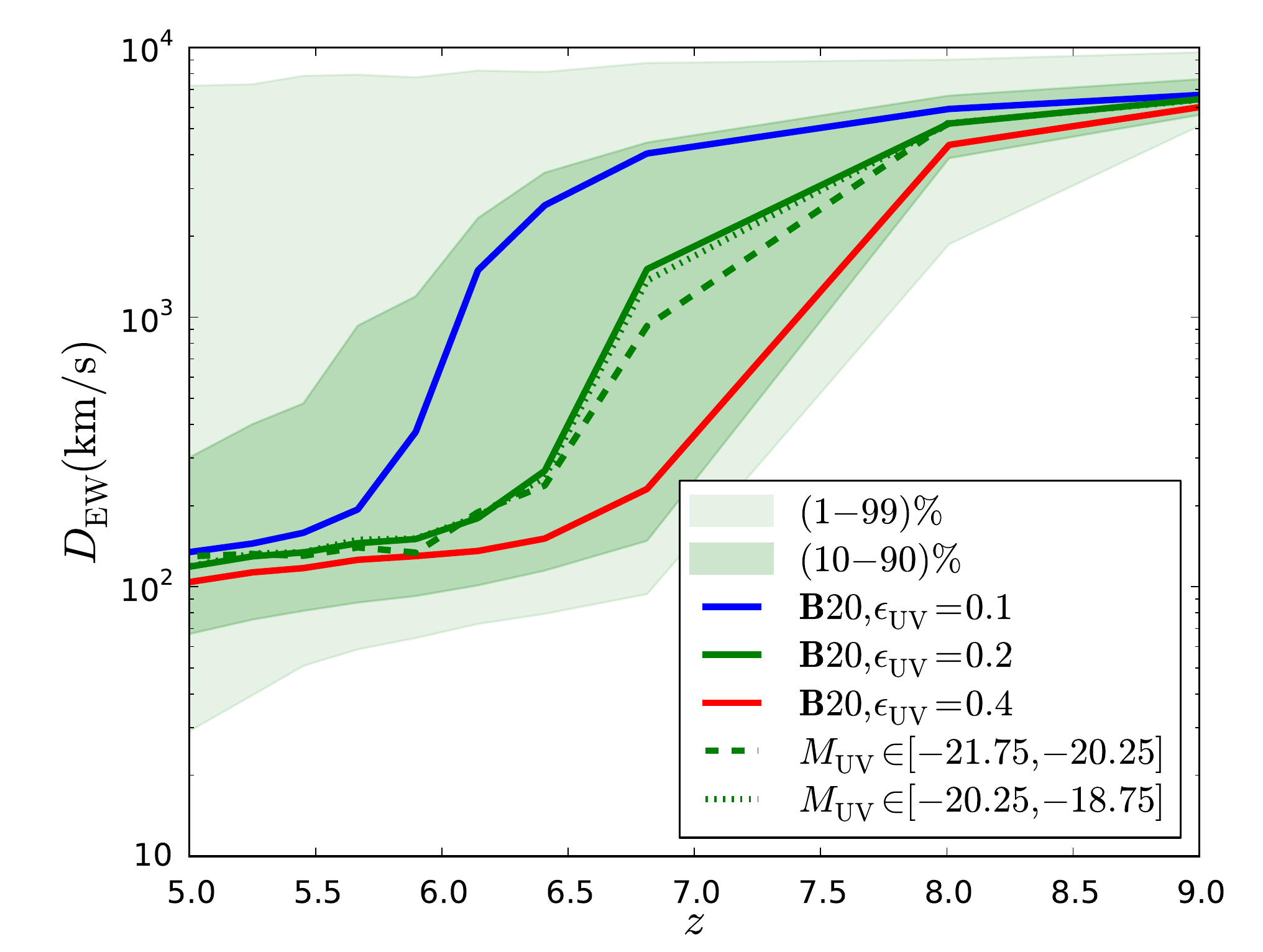}
\includegraphics[width=\hsize]{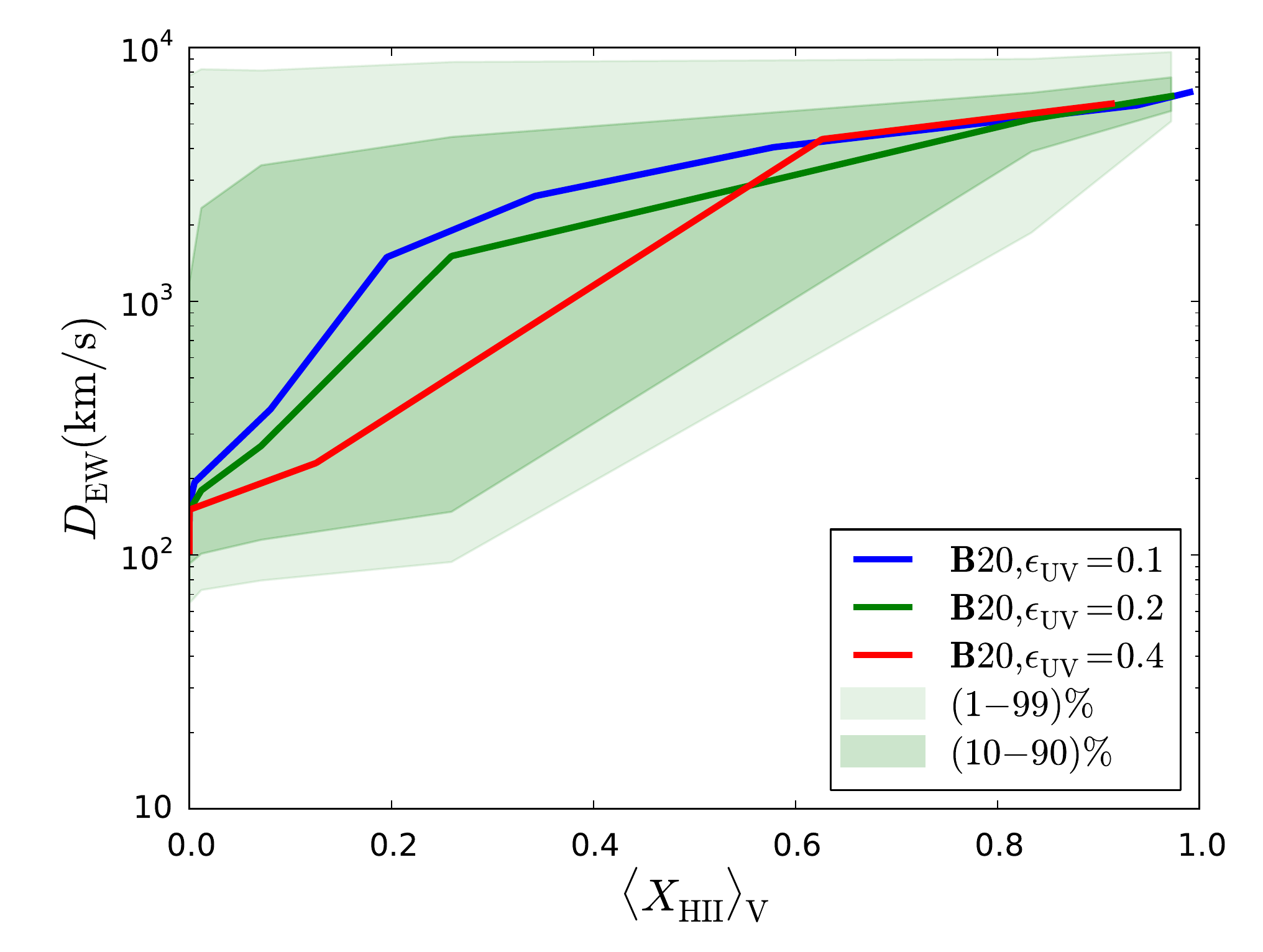}
\caption{Equivalent width of \lya\ absorption $\dew$ as a function of redshift (top) or volume-weighted neutral hydrogen fraction (bottom). Colored lines show averages over all realizations of simulation sets B20.uv4 (red), B20.uv2 (green), and B20.uv1 (blue) for all galaxies with UV magnitudes between -22 and -18. Green semi-transparent bands give the distribution of $\dew$ around the mean (as 1-99 and 10-90 percentiles). Green dashed and dotted lines show $\dew$ for the subsets of galaxies with magnitudes in bins [-21.72,-20.25] and [-20.25,-18.75] respectively.\label{fig:dwing}}
\end{figure}

From comparison between Figures \ref{fig:xhz} and \ref{fig:dwing} it is clear that beginning of the overlap of ionized bubbles corresponds to a rapid decrease in the equivalent width of the damping wing - in our fiducial B20.uv2 run $\dew$ drops by an order of magnitude between $z=7$ and $z=6.5$. Such behavior is consistent with the observed sharp decline in the fraction of \lya\ emitters between $z=6$ and $z=7$ \citep{lae:pfv11,lae:sse12,lae:cbw14}, although, as we mentioned above, a more quantitative comparison would require a better model of \lya\ emitters in the simulations.

We also notice that in our simulations there is little dependence of the damping wing equivalent width $\dew$ on galaxy luminosity - dotted and dashed lines in Figure \ref{fig:dwing} show the evolution of $\dew$ for two luminosity bins, but both lines trace similar behavior. The luminosity dependence of the fraction of \lya\ emitters has been seen in some observational studies \citep[c.f.][]{lae:sse12} but not in others \citep[c.f.][]{lae:pfv11}. If such dependence is confirmed by further observations, it would imply that brighter galaxies have intrinsically higher probability of becoming a \lya\ emitter than fainter ones.

The observed measurements of the fraction of galaxies that remain strong \lya\ emitters have been also used to constraint the mean volume-weighted neutral fraction. The bottom panel of Figure \ref{fig:dwing} replaces the redshift axis with the (monotonically decreasing with time) volume-weighted $\HI$ fraction. The sensitivity of $\dew(\langle X_\HI\rangle_V)$ to the $\euv$ parameter is much less than when $\dew$ is treated as a function of $z$, confirming the validity of the assumption that the decrease in the observed fraction of \lya\ emitter at higher redshifts indicates a change of the volume-weighted average neutral fraction. In fact, it appears that a condition $\dew<1000\dim{km/s}$ corresponds to $\langle X_\HI\rangle_V\la0.2$, while a condition $\dew<500\dim{km/s}$ corresponds to $\langle X_\HI\rangle_V\la0.1$. This conclusion is in good agreement with other recent simulation studies \citep[c.f.][]{lae:tl14,newrei:hdp14}.

\subsection{Back Reaction of Reionization on Early Galaxies}
\label{sec:back}

\begin{figure}[t]
\includegraphics[width=\hsize]{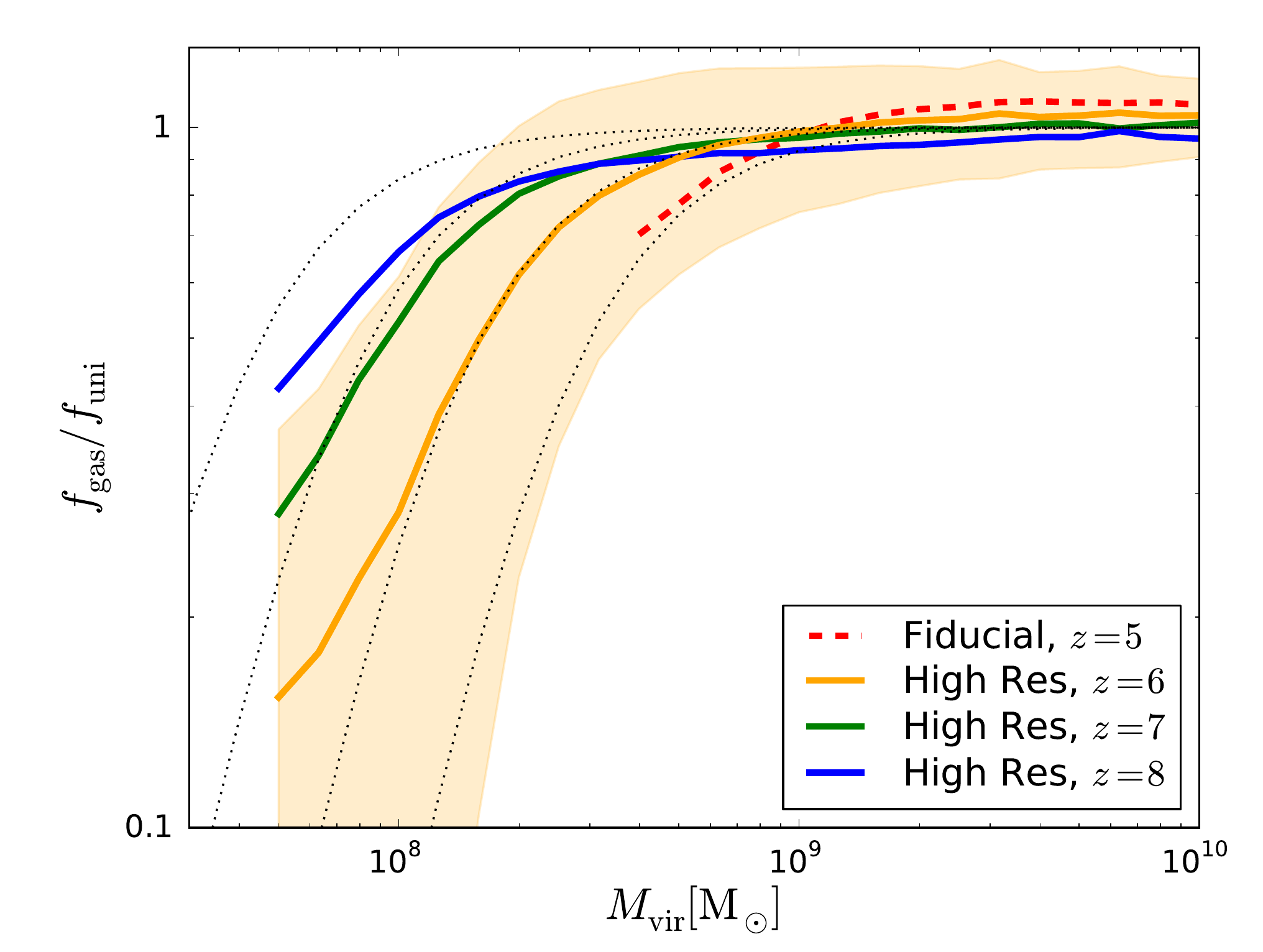}
\caption{Average gas fractions (in units of the universal) for the high resolution run B20HR.uv2 at $z=6$, $7$, and $8$ (solid lines) and for the fiducial B20.uv2 run at $z=5$ (at earlier times the fiducial run lack enough mass resolution). A light orange band shows the rms scatter around the $z=6$ line for the B20HR.uv2 run. Thin black lines are fits from \citet{dsh:ogt08}, that agree with our simulations after reionization ($z\la6$) almost perfectly.\label{fig:fgas}}
\end{figure}

The most immediate effect of reionization on the early galaxies - or, rather, galactic halos - is to expel photoionized gas from sufficiently low mass halos. This process, sometimes inaccurately called ``photoevaporation'', has been a focus of a large number of studies, reviewing which is beyond the scope of this paper. So far, the highest mass resolution (up to $3\times10^4\Msun$ - a critical numerical parameter for this question) has been achieved by \citet[][they also review the previous works]{dsh:ogt08}, who provided accurate fits for the average gas fraction as a function of halo mass and redshift. In Figure \ref{fig:fgas} we compare the gas fractions in our simulations with the fits from \citet{dsh:ogt08}. Our fiducial simulation sets barely have enough mass resolution to properly capture the characteristic mass below which halos start losing gas due to photoionization. A high resolution run B20HR.uv2 captures that effect well, and our results post-reionization (at $z\la6$) agree with \citet{dsh:ogt08} fits extremely well. At earlier redshifts the difference is expected, as \citet{dsh:ogt08} assumed an instantaneous reionization at $z=9$, while we model the actual reionization history.

A good agreement of our results with \citet{dsh:ogt08} illustrates a simple, but not widely appreciated fact that it takes only a few hundred million years for the effect of reionization on the gas fractions to be fully established \citep{rei:isr05} - for example, our B20HR.uv2 run reionizes at $z\approx7.3$, and fully converges with the \citet{dsh:ogt08} simulations by $z=6$, only $210\dim{Myr}$ later.

\begin{figure}[t]
\includegraphics[width=\hsize]{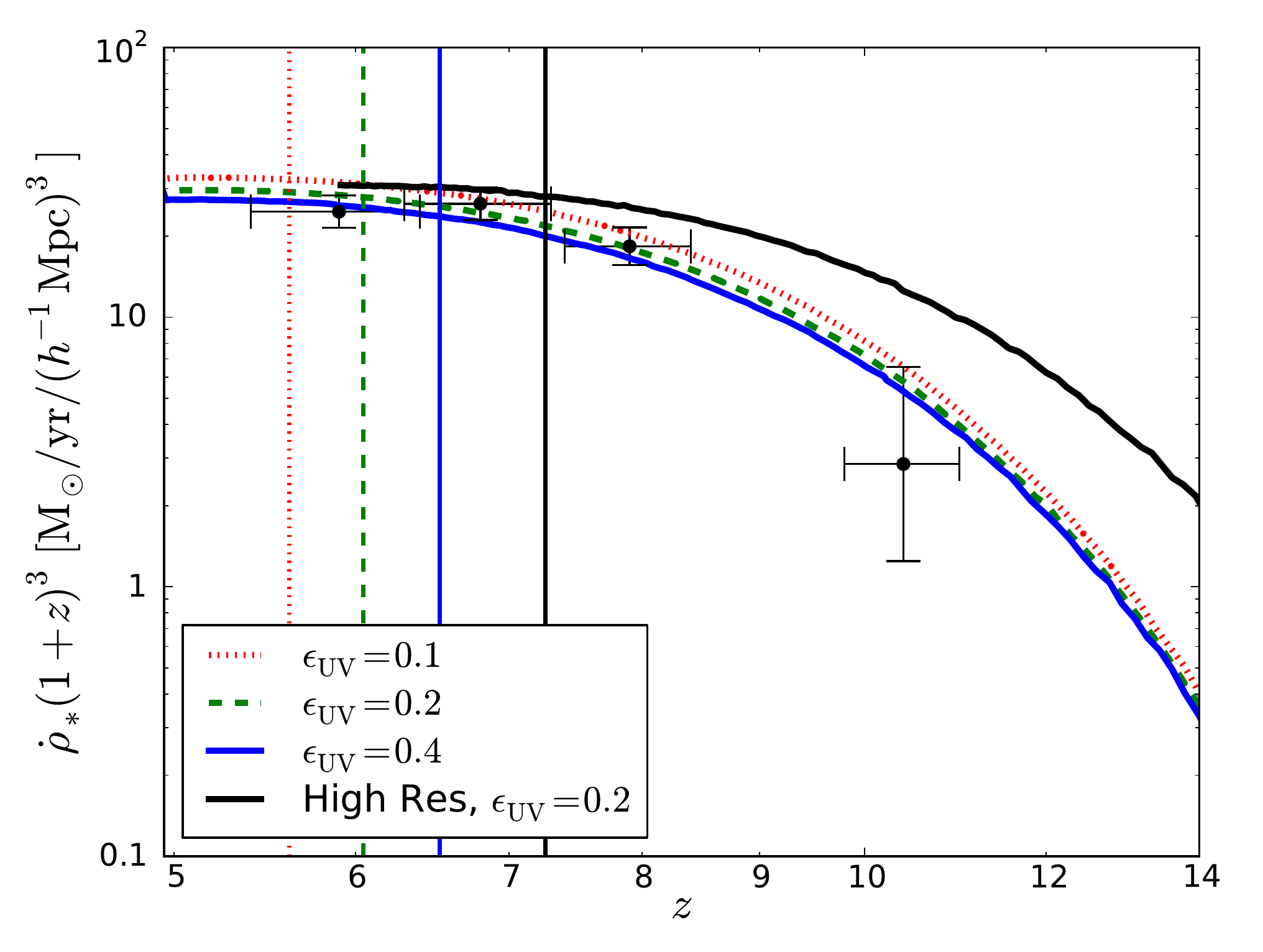}
\caption{Cosmic star formation histories for the three fiducial simulation sets with varied ionizing intensity and the high resolution run B20HR.uv2. Thin vertical lines mark reionization redshifts for each simulation set (defined as times when the volume weighted neutral fraction falls below $10^{-3}$, see \protect\citet{ng:g00a}). Black points with error-bars are the data from \protect\citet{rei:biot14}. Reionization does not introduce any noticeable feature in the global star formation history.\label{fig:sfrz}}
\end{figure}

A secondary effect of reionization is on the actual star formation rates in early galaxies. That effect can be large or nil, depending on the fraction of star formation in halos that are affected by photoionization. The simplest manifestation of such back reaction is a change in the global cosmic star formation history. Studies of the response of the global star formation history to reionization were pioneered by \citet{reiback:bl00}, who found a large suppression in the global star formation history at reionization. This ``Barkana \& Loeb'' effect has been revisited repeatedly in the previous studies \citep{reiback:tab03,reiback:wl06,reiback:bl06,reiback:dfo06,reiback:wc07,reiback:pm07,reiback:yok07,reiback:ml11,reiback:dwm14}, with different groups disagreeing significantly in its strength. Global star formation histories of our three fiducial simulation sets as well as the high resolution run are B20HR.uv2 shown in Figure \ref{fig:lfend}. No effect of reionization is noticeable in the figure, in contradiction with some of the previous studies. 

\begin{figure}[t]
\includegraphics[width=\hsize]{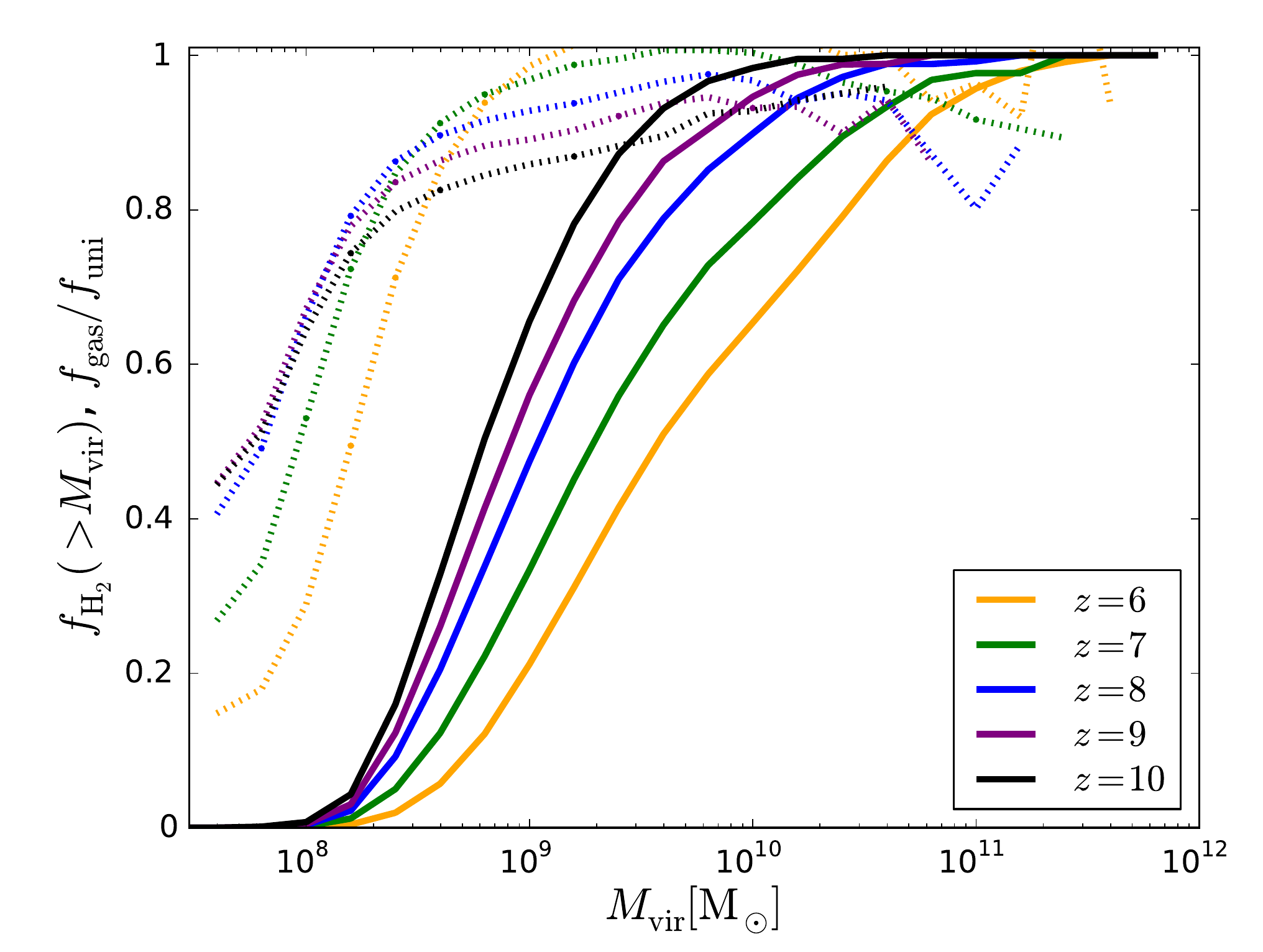}
\caption{Average gas fractions (dotted lines - the same as shown in Fig.\ \protect\ref{fig:fgas}) and cumulative molecular gas fractions in halos above a certain mass (solid lines) for the high resolution run B20HR.uv2 at several redshifts. Notice that halos affected by reionization contain little molecular gas and, hence, form no stars. \label{fig:fh2}}
\end{figure}

In order to elucidate that disagreement, we show in Figure \ref{fig:fh2} the cumulative fraction of molecular gas in halos above a given mass at several redshifts. There is always at least a decade in mass difference between the halos that are affected by reionization and halos that contain substantial amount of molecular gas. Since our model for star formation is crucially based on the (observationally motivated) paradigm that stars form primarily in the molecular gas, the negligible back reaction on the star formation in early galaxies is naturally explained by the large difference in the two mass scales.

As a side note, we recall that our fiducial runs and the high resolution run B20HR.uv2 both reproduce observed galaxy UV luminosity functions at all redshifts $z\ga6$ (Figure 8 of Paper I), but they have global star formation histories that differ at $z\approx10$ (Fig.\ \ref{fig:sfrz}) by more than the claimed observational errorbars from \citet{rei:biot14}. Hence, we conclude that the observational determination of the global star formation history is based on the assumptions that do not always hold, and, hence, the systematic errors of such determinations are substantially larger than the formal statistical errors at $z>8$.

\begin{figure}[t]
\includegraphics[width=\hsize]{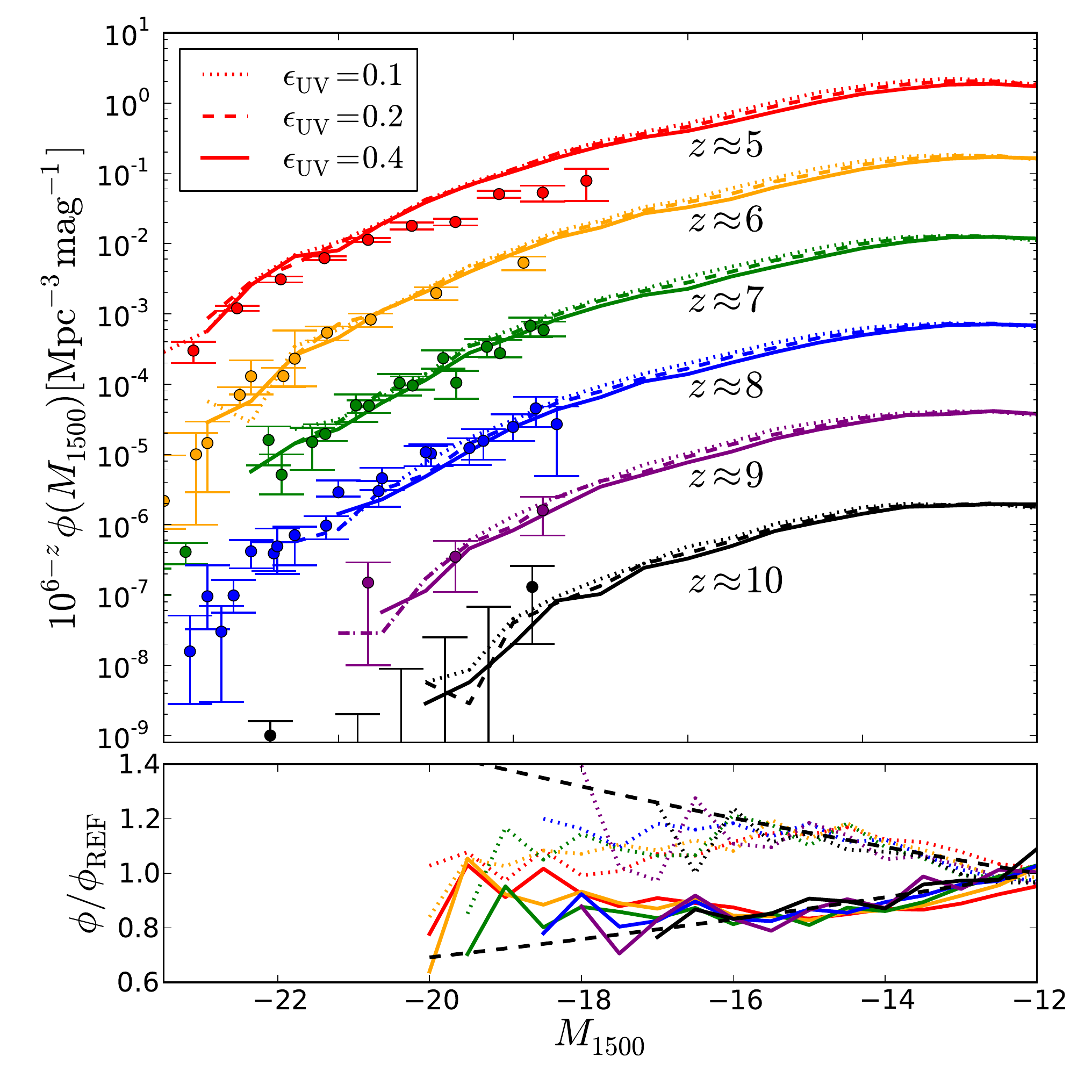}
\caption{Top: ultraviolet galaxy luminosity functions for the three simulation sets with varied ionizing intensity at 6 different redshifts (as shown in the legend). Circles with error-bars are a compilation of recent observational measurements \citep{gals:biff07, gals:biol11,rei:obig12,rei:btos12,rei:sreo13,rei:wmhb13,rei:obil13,rei:bdm14,rei:obi14}. Different redshifts are shifted vertically by 1 dex for clarity. Bottom: differences between simulation sets with $\euv=0.1$ (dotted lines) and $\euv=0.4$ (solid lines) relative to the reference value of $\euv=0.2$. Black dashed lines show the variation in the faint-end slope by $\pm0.05$.\label{fig:lfend}}
\end{figure}

While not affecting the global star formation history in any significant way, reionization may still leave more subtle signatures in the properties of early galaxies. For example, the sensitivity of the faint end slope of the galaxy luminosity function to the reionization history has been proposed as an important science goal for JWST \citep{misc:gmcd06}. To explore the feasibility of such a test, we show in Figure \ref{fig:lfend} galaxy luminosity functions for the three fiducial simulation sets with varying ionizing emissivity. In this case the effect of reionization is detectable, although it is not large. As the bottom panel of Fig.\ \ref{fig:lfend} shows, a factor of two variation in the ionizing emissivity (which corresponds to about $\Delta z\approx0.5$ change in the redshift of reionization - see Fig.\ \ref{fig:dwing} and Fig.\ 4 of Paper I) from our fiducial value $\euv=0.2$ produces a change of about 0.05 in the faint end slope of the luminosity function for $M_{1500} \ga -17$, although at brighter magnitudes the difference rapidly disappears. This variation is likely to be too small to be usable as a constraint on the reionization history.

\section{Conclusions}
\label{sec:con}

We present reionization history and global characteristics of the reionization process from a suite of recent numerical simulations performed as part of the Cosmic Reionization On Computers (CROC) project. CROC simulations reproduce the observed evolution of the galaxy UV luminosity function between $z=10$ and $z=6$ well, and, hence, include realistic treatment of the dominant class of ionizing sources. 

We find that, in order to match the observational constraints on the post-reionization \lya\ forest at $5<z<6$, we need to set the ionizing emissivity parameter $\euv$ (that measures the escape fraction up to the resolution limit of our simulations) to just under $\euv=0.2$. However, as we also emphasized in Paper I, cosmic variance increases sharply with redshift, and at $z>6$ our simulations do not yet converge on the global properties of the IGM, such as the mean Gunn-Peterson optical depth or the volume weighted $\HI$ fraction. Since the statistical power of our simulations is much higher than the statistical reach of the existing absorption spectra of high-redshift quasars, we conclude that, unfortunately, the observations are unlikely to have reached the convergence either.

In a further illustration of this, we show that the distribution of the Gunn-Peterson optical depth over the redshift intervals $\Delta z\approx 0.15$ is extraordinary wide at $z>6$, but even at $z<6$ the $\tau_{\rm GP}$ distribution retains a relatively long tail towards high values.

The distributions of ionized and neutral bubbles during most of cosmic reionization is approximately flat, meaning that it is roughly equally likely for a random place of the universe to be in a large or a small bubble. We find good numerical convergence in bubble sizes down to $z\sim7$, at which point the finite sizes of our simulation boxes start biasing the distribution of ionized bubbles. That result illustrates the importance of achieving consistent numerical resolution between the gas dynamic solver and the radiative transfer solver - the mismatch between the two resolution likely results in erroneous over-propagation of ionizing radiation beyond the few mean free path lengths.

We show that the equivalent width of the damping wing of \lya\ absorption increases rapidly from mellow values of $\dew\sim100\dim{km/s}$ at $z=6$ to whopping $\dew\sim2000\dim{km/s}$ by $z=7$. While $\dew$ serves only as a rough proxy for the suppression of galaxy \lya\ emission line by the neutral IGM in front of it, this result is generally consistent with the observed sharp decline in the fraction of \lya\ emitting galaxies at $z=7$ as compared to $z=6$. We also confirm conclusions from the previous simulation and analytical work that such suppression corresponds to substantial, but not dominant volume weighted neutral fraction of about 0.2.

While our results on the reionization history are in good agreement with most of prior studies, we find little back reaction of reionization on the properties of early galaxies. Because galaxies that are affected by photoionization contain little molecular gas (and, hence, star formation), we find that the global star formation history is insensitive to the reionization history, i.e.\ the ``Barkana \& Loeb'' effect does not exist. A more subtle effect of reionization is in modifying the faint end slope of the galaxy UV luminosity function, but such a modification is rather small (change in the slope of about 0.1 for a unit shift in the redshift of reionization). Since predicting the faint end slope to such precision theoretically would be extremely challenging, we conclude that, unfortunately, measuring the faint end slope by JWST will not be a useful constraint on reionization, contrary to expectations \citep{misc:gmc06}.

One observational constraint that we have ignored so far is the optical depth to Thompson scattering from the CMB observations by the \emph{WMAP} mission. While the history of \emph{WMAP} measurements of the Thompson optical depth is rocky, the latest value from the 9-year \emph{WMAP} data is $0.089\pm0.014$ \citep[or $0.081\pm0.012$ if other data are included in a joint fit,][]{wmap9a,wmap9b}. The value we get for fiducial sets B20.uv2 and B40.uv1 is $0.052\pm0.003$, and in the set B40.uv2 that reionizes earlier, the value for the Thompson optical depth only rises to $0.057\pm0.004$, which are only marginally (at $2\sigma$ level) consistent with the \emph{WMAP} values.

A large portion, if not all, of this discrepancy is due to incomplete numerical convergence of our simulations. In Paper I we compared our fiducial runs (an equivalent of $512^3$ particles in a $20h^{-1}\dim{Mpc}$ box) with a single higher mass resolution run B20HR.uv2 that we were able to complete (an equivalent of $1024^3$ particles in a $20h^{-1}\dim{Mpc}$ box). While numerical converge tests indicate that our fiducial runs account for 55\% of all ionizing photons, the higher reslution  B20HR.uv2 run accounts for 80\% of them. As the result, the Thompson optical depth raises to 0.067 in that run. Simple linear extrapolation to the limit of 100\% of ionizing radiation gives a value of 0.08 for the Thompson optical depth, fully consistent with the current observational measurements. 

Whether incomplete numerical convergence is, indeed, a full story will have to wait for more powerful computers, however, as at present we are unable to run the whole ensemble of higher mass resolution simulations - for example, a higher mass resolution equivalent of our planned $80h^{-1}\dim{Mpc}$ run would have $4096^3$ particles and will require of order of 200 million CPU hours, the amount not currently feasible to obtain for this kind of work.

\acknowledgements

We are grateful to George Becker for valuable comments on the early draft of this paper.

Simulations used in this work have been performed on the Joint Fermilab - KICP cluster ``Fulla'' at Fermilab, on the University of Chicago Research Computing Center cluster ``Midway'', and on National Energy Research Supercomputing Center (NERSC) supercomputers ``Hopper'' and ``Edison''.

\bibliographystyle{apj}
\bibliography{ng-bibs/self,ng-bibs/ism,ng-bibs/misc,ng-bibs/sims,ng-bibs/sfr,ng-bibs/rei,ng-bibs/dsh,ng-bibs/qlf,ng-bibs/gals,ng-bibs/igm,ng-bibs/reisam,ng-bibs/lae,ng-bibs/newrei,ng-bibs/reiback,ng-bibs/cosmo}

\end{document}